# Association of intermittency with electron heating in the near-Sun solar wind

C. Phillips[1], R. Bandyopadhyay[1], D. J. McComas[1], S. D. Bale[2,3,4,5]

[1]Department of Astrophysical Sciences, Princeton University, Princeton, NJ 08544, USA
[2]Space Sciences Laboratory, University of California, Berkeley, CA 94720-7450, USA
[3]Physics Department, University of California, Berkeley, CA 94720-7300, USA
[4]The Blackett Laboratory, Imperial College London, London, SW7 2AZ, UK
[5]School of Physics and Astronomy, Queen Mary University of London, London E1 4NS, UK

Several studies in the near-Earth environment show that intermittent structures are important sites of energy dissipation and particle energization. Recent Parker Solar Probe (PSP) data, sampled in the near-Sun environment, have shown that proton heating is concentrated near coherent structures, suggesting local heating of protons by turbulent cascade in this region. However, whether electrons exhibit similar behavior in the near-Sun environment is not clear. Here, we address this question using PSP data collected near the Sun during the first seven orbits. We use the partial variance of increments (PVI) technique to identify coherent structures. We find that electron temperature is preferentially enhanced near strong discontinuities, although the association is somewhat weaker than that with protons. Our results provide strong support for inhomogeneous heating of electrons in the "young" solar wind, associated with dissipation of turbulent fluctuations near intermittent structures.

**Introduction**
Most space plasmas are in turbulent states. Dissipation of the turbulent fluctuations into heat has important effects on space plasmas (Matthaeus & Velli 2011). Turbulent heating may be responsible for the super-adiabatic temperature profile observed in the solar wind (Richardson et al. 1995) and for solar wind acceleration. Turbulent heating may also provide an explanation of the coronal heating problem. Several previous works have established that an energy cascade exists in the solar wind at length scales much larger than the ion kinetic scales (e.g., MacBride et al. 2005; Sorriso-Valvo et al. 2007). However, the details of how this energy is dissipated and how different charged species – ions and electrons – are heated in a turbulent plasma are still not well understood, although many possible mechanisms have been proposed. For a review of possible heating mechanisms responsible for the coronal heating and solar wind acceleration, see McComas et al. (2007).

A leading candidate is heating at current sheets and other coherent structures that are abundantly present in turbulent plasmas (Matthaeus et al. 2015). These current sheets are possible sites of magnetic reconnection which may contribute significantly to the overall heating. Regardless of the mechanism, several previous studies, both numerical and observational, have shown that protons and electrons are heated preferentially near current sheets (Chasapis et al. 2015; Osman et al. 2010a; Parashar & Matthaeus 2016) in highly inhomogeneous, localized regions. Additionally, temperature anisotropy, energetic particle flux, and kinetic microinstabilities are also observed to be enhanced near the current sheets (Osman et al. 2012; Tessein et al. 2013; Bandyopadhyay et al. 2020; Qudsi et al. 2020). Therefore, even though they occupy only a small fraction of the total volume, these coherent structures play an active role in the dynamics of solar wind and other turbulent plasmas (Yordanova et al. 2021).

NASA's Parker Solar Probe (PSP) provides the first opportunity to study the intermittent structures in the near-Sun solar wind environment and investigate their relationship to dissipative processes (Fox et al. 2016). A main goal of the PSP mission is to shed light into the processes responsible for heating of the lower corona. Recent PSP observations in the near-Sun environment have shown statistical correlation between coherent magnetic field structures and enhanced proton temperature, suggesting heating by reconnection at current sheets (Qudsi et al. 2020; Sioulas et al. 2022). However, a quantitative observational assessment of electron heating in thin current sheets has not yet been reported in the near-Sun environment. Here we examine whether a similar association of intermittency and heating is seen in the near-Sun solar wind electrons.

**Data & Methodology**
A practical technique for identifying the presence of coherent structures is the method of Partial Variance of Increments (PVI). Following Greco et al. 2008, PVI of the magnetic field ($\boldsymbol{B}$) at time $t$ is defined as

$$\text{PVI}(t, \tau) = \frac{|\Delta \boldsymbol{B}(t, \tau)|}{\sqrt{\langle |\Delta \boldsymbol{B}(t, \tau)|^2 \rangle}} \qquad (1)$$

Where $\tau$ is a particular time lag, $\Delta \boldsymbol{B}$ is an increment of the magnetic field defined as $\Delta \boldsymbol{B} = \boldsymbol{B}(t + \tau) - \boldsymbol{B}(t)$, and $\langle ... \rangle$ is a time average over a reasonably large span – larger than the correlation time. The PVI is essentially the strength the field fluctuation at a given scale (calculated by increment), normalized by the variance of the field. The PVI is closely related to the structure function of first order, but unlike structure function, it is a pointwise measure rather than an average. Events with PVI values exceeding 3 are associated with super-Gaussian structures, such as current sheets. We note that there have been several other more complicated methods developed for identifying gradients in turbulent flows (see Greco et al. 2017 for a review), such as the Tsurutani-Smith method (Tsurutani & Smith 1979), wavelet-based local-Intermittency Measure (Veltri & Mangeney 1999; Farge et al. 2001), and Phase Coherence Index (Hada et al. 2003). Here, we use the PVI method for its simplicity.

One of the main outstanding problems in collisionless plasmas is that, unlike in magnetohydrodynamics (MHD), the dissipation function is not known from the first principles (Chapman & Cowling 1990). Recent works have proposed several quantitative measures of dissipation in collisionless plasmas, but most of those cannot be evaluated from single-spacecraft data (e.g., Cerri 2016; Yang et al. 2017; Chen et al. 2020; Argall et al. 2022). Therefore, we use the electron temperature as a tentative proxy for local electron heating, as has been done in previous works (Chasapis et al. 2015; Osman et al. 2010a). Enhanced temperature is not identical to heating, but it may be viewed roughly as a proxy if the heat conduction is not too strong (see Osman et al. 2010a for a detailed discussion). We use electron temperature and magnetic field data from PSP's first 7 encounters with the Sun. The latter are averaged to the cadence of the former for our analyses. The magnetic field data are collected by the fluxgate magnetometer (MAG) from the FIELDS (Bale et al. 2016, 2019). Electron temperature data were obtained from the quasi-thermal noise (QTN) spectrum provided by the electric field antennas in the FIELDS suite (Pulupa et al. 2017; Moncuquet et al. 2020; Martinović et al. 2022). This provides the core electron temperature at a cadence of 6.99 s close to the perihelia.

To calculate PVI according to Equation (1), we use magnetic field data from the fluxgate magnetometer (MAG) instrument onboard PSP. We use a lag of $\tau = 1.195$ NYs (1 s). Using the Taylor hypothesis (Taylor 1938) with a typical 300 km/s solar wind near the sun, this lag corresponds to a length scale of roughly 300 km. The correlation length is about ~$10^5$ km and the ion-inertial length is about 15 km (Chen et al. 2020; Parashar et al. 2020). So, the lag of ~300 km in within the turbulence inertial range. The averaging is performed over 2 hours, which is several times larger than the correlation time. The PVI time series was then resampled to the electron temperature cadence such that for each interval of 6.99 second. The maximum value of PVI in that interval was chosen as current sheets can be very narrow structures. In this study, we focus on the near-Sun environment at radial distances in the range 0.09 AU < r < 0.18 AU. However, performing the same analyses at 0.18 AU < r < 0.25 AU gives qualitatively the same results.

**Results**

We begin by showing an example of coherent structures with high PVI values, along with other relevant quantities, in Fig. 1. The magnetic field magnitude, RTN components, electron temperature, and PVI from a period in encounter 1 is shown, where vertical lines mark points with PVI > 3. We see that, like the coherent structures, the high temperature events are also clustered together in an intermittent fashion (Chhiber et al. 2020). In the following, we investigate the relation of the electron temperatures with intermittent structures statistically.

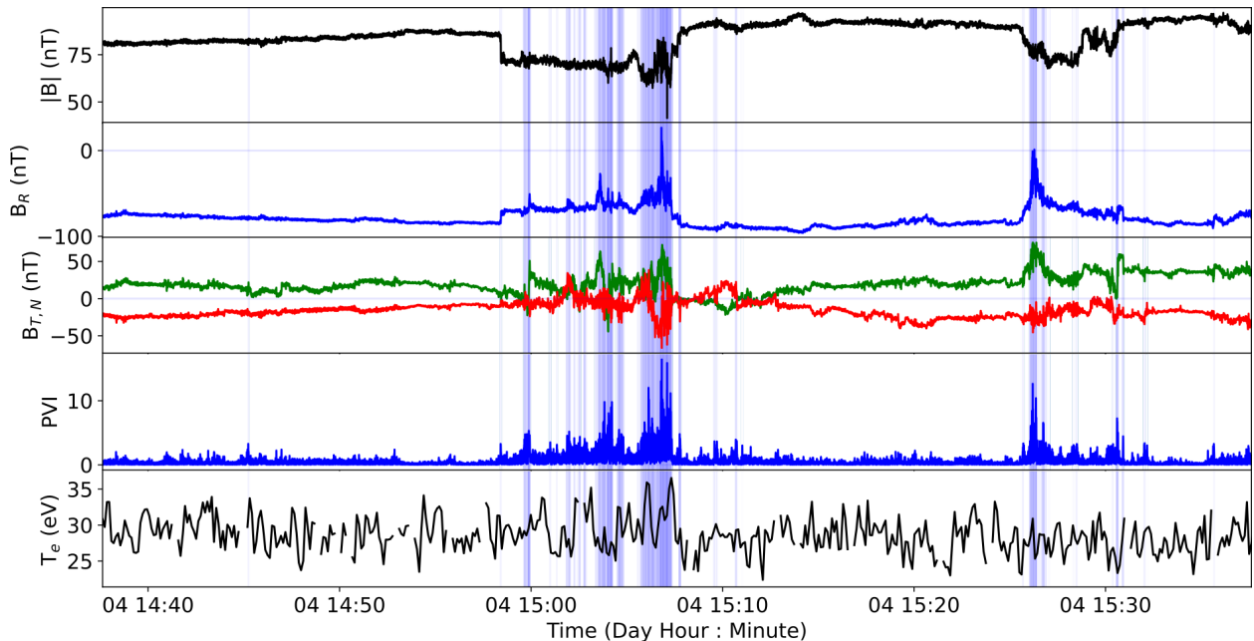

**Figure 1.** An example of clustering of intermittent structures associated with elevated electron temperature. From top, the magnitude of the magnetic field |**B**|, the radial component of the magnetic field ($B_R$), the tangential component of the magnetic field ($B_T$) in green and normal ($B_N$) component in red, the magnetic-field PVI, and the electron temperature $T_e$ are shown. The shaded regions represent the locations of PVI > 3 events.

Figure 2 shows the electron temperature versus PVI for r < 0.18 AU, averaged over all 7 orbits. Each bin contains an equal number of data points. Figure 2 shows a clear positive trend, where the average electron temperature increases with the PVI values. This statistical correlation suggests that electron heating processes are active at high PVI regions.

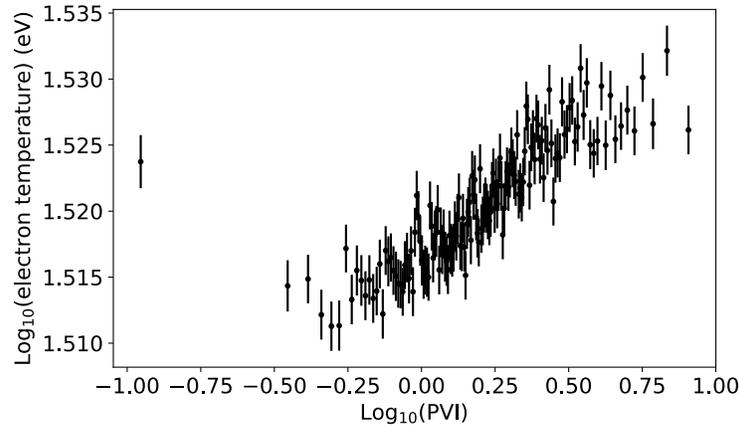

**Figure 2.** Logarithm of average electron temperature value for each PVI bin for PSP data close to the Sun. The stronger PVI events are associated with the higher values of temperature.

Next, we gather the electron temperature data in bins of increasing values of PVI and divide all the data points into 6 bins such that each bin has equal number of points. We then calculate the temperature distribution within each bin, as shown in Fig. 3. The vertical lines are the mean values for their respective color. There is a monotonic increase in average temperature with increasing PVI, where the lowest PVI bin has a mean temperature of 33.4 eV and the highest PVI bin has a mean temperature of 34.4 eV. While small, this is a statistically significant change in the electron temperature with increasing PVI. This result further supports that electron heating occurs close to the coherent structures, represented by high PVI.

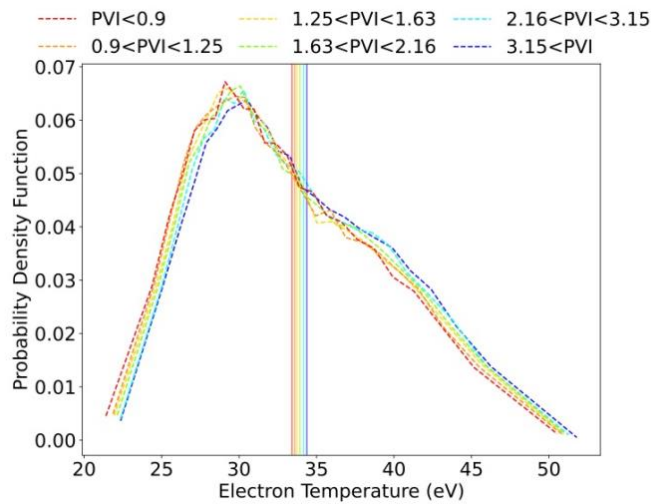

**Figure 3.** Probability distribution functions (PDFs) of the electron temperature corresponding to different PVI ranges. The probability density increases with the increase in temperature for high PVI, while it decreases in temperature for low PVI. The Vertical lines with different colors show the mean electron temperature for each corresponding PDF.

To further examine the relationship between intermittent structures and electron heating, we study the electron temperature in the vicinity of high PVI events using the methodology described by (Osman et al. 2010a, 2012). For a selected PVI threshold, we compute the average electron temperature at the PVI event and for points before and after it, up to about one correlation length (equivalent to approximately 10 min). Formally, these averages may be quantified as

$$\widetilde{T}_e(\Delta t, \theta_1, \theta_2) = \langle T_e(t_{PVI} + \Delta t) | \theta_1 \leq PVI \leq \theta_2 \rangle,$$

where $\widetilde{T}_e$ is the conditionally average temperature, $t$ is the time of a particular PVI event satisfying the criteria that its value is within the thresholds $\theta_1$ and $\theta_2$, and $\Delta t$ is the time from the PVI events.

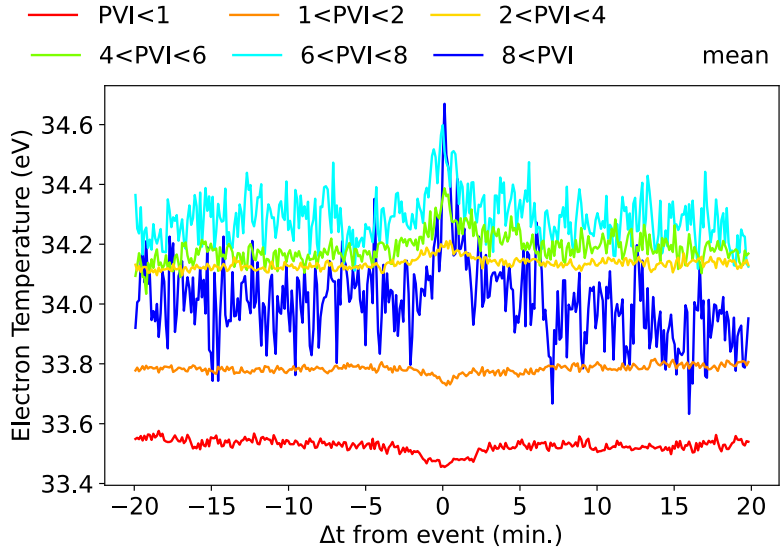

**Figure 4.** Conditional average of electron temperature for different PVI thresholds. The electron temperature is elevated at the instant of PVI event and continues to have an elevated temperature in its vicinity. The red and orange curves, corresponding to the low PVI, show a depression at the origin, suggesting absence of heating when the magnetic field is very smooth.

Figure 4 plots the electron temperature for various thresholds of PVI values. Consistent with Figs. 2 & 3, we observe an elevated electron temperature at the point of high PVI events. Additionally, the electrons temperature values are most strongly increased near the PVI events and decrease within minutes away from these events. There is a characteristic core of elevated temperature with half width of approximately 2.5 min ($\equiv$ 45000 km, assuming a 300 km/s solar wind speed) and the profiles become approximately flat at distances beyond a correlation scale ($\sim 1.5 \times 10^5$ km $\equiv$ 500 s). This indicates that the electrons are preferentially heated in the general vicinity of coherent structures.

Another feature noticeable in Fig. 4 is the depression in the temperature profile at the origin for the small PVI ranges. Contrary to the high-PVI cases, the electron temperature values gradually increase away from the origin. The low PVI regions indicate where the magnetic field is smooth, and the plots in Fig. 4 show that in these regions the electron temperature is lower than the surrounding plasma. This is consistent with weaker heating of electrons in these regions.

**Discussion & Conclusions**

The present work is the first statistical study of electron heating in current sheets in the near-Sun environment. Using PSP data from the first seven orbits, collected near the sun, we study the association of electron temperature with intermittent structures. We find that electron temperatures are well correlated with high-PVI events. Further, we find that the electrons are generally hotter near the coherent structures, identified by high PVI values. Of course, elevated temperature is not identical to local heating, but in absence of a dissipation function we use the electron temperature as a tentative proxy. Similar results were obtained for protons by Qudsi et al. 2020. Our results are also consistent with the findings in other studies which examine these effects in the near-Earth solar wind (Osman et al. 2010b, 2012). Our results support the proposition that significant inhomogeneous electron heating occurs in the "nascent" solar wind, connected with current sheets dynamically generated by turbulence.

Here, we have used the QTN-spectroscopy based electron temperature which gives the temperature of the core of the distribution. The different components of the electron distribution function – the core, the strahl, and the halo – may provide insights into how the different populations behave near the current sheets. The behavior of electron temperature anisotropy around intermittent structures also remains to be studied.

Further, late last year PSP began sampling the sub-Alfvénic coronal wind, inside the Alfvén critical surface (Bandyopadhyay et al. 2022; Kasper et al. 2021; Zank et al. 2022). Data from these intervals and future orbits will reveal whether similar behavior persists in the sub-Alfvénic coronal wind, which will provide more insights into the nature of turbulent heating processes occurring in the lower corona.

In summary, our results indicate that turbulent dissipation already begins contributing to electron heating near the PSP perihelia, inside ~0.2 au. This result provides important new information for modeling heating in the corona and solar wind.

**Acknowledgements**

We are deeply indebted to everyone that helped make the Parker Solar Probe (PSP) mission possible. This work was supported as a part of the PSP mission under contract NNN06AA01C. This research was partially supported by the Parker Solar Probe project through Princeton/ISʘIS subcontract SUB0000165 and in part by PSP GI grant 80NSSC21K1767 at Princeton University. The Parker Solar Probe was designed, built, and is now operated by the Johns Hopkins Applied Physics Laboratory as part of NASA's Living with a Star (LWS) program (contract NNN06AA01C). Support from the LWS management and technical team has played a critical role in the success of the Parker Solar Probe mission. All the data, used in this paper, are publicly available via the NASA Space Physics Data Facility (https://spdf.gsfc.nasa.gov/).

**References**

Argall, M. R., Barbhuiya, M. H., Cassak, P. A., et al. 2022, Phys Plasmas, 29 (American Institute of Physics), 022902

Bale, S., Badman, S., Bonnell, J., et al. 2019, Nature (Nature Publishing Group), 1, https://doi.org/10.1038/s41586-019-1818-7

Bale, S. D., Goetz, K., Harvey, P. R., et al. 2016, Space Sci Rev, 204, 49, https://doi.org/10.1007/s11214-016-0244-5
Bandyopadhyay, R., Goldstein, M. L., Maruca, B. A., et al. 2020, Astrophys J Suppl Ser, 246, 48
Bandyopadhyay, R., Matthaeus, W. H., McComas, D. J., et al. 2022, Astrophys J Lett, 926 (American Astronomical Society), L1, https://doi.org/10.3847/2041-8213/ac4a5c
Cerri, S. S. 2016, Plasma turbulence in the dissipation range - theory and simulations (Universität Ulm und Technischen Hochschule), http://dx.doi.org/10.18725/OPARU-3355
Chapman, S., & Cowling, T. G. 1990, The mathematical theory of non-uniform gases: an account of the kinetic theory of viscosity, thermal conduction and diffusion in gases (Cambridge university press)
Chasapis, A., Retino, A., Sahraoui, F., et al. 2015, Astrophys J Lett, 804, L1, http://doi.org/10.1088/2041-8205/804/1/L1
Chen, C. H. K., Bale, S. D., Bonnell, J. W., et al. 2020, Astrophys J Suppl Ser, 246 (American Astronomical Society), 53, https://doi.org/10.3847/1538-4365/ab60a3
Chhiber, R., Goldstein, M. L., Maruca, B. A., et al. 2020, Astrophys J Suppl Ser, 246 (American Astronomical Society), 31, https://doi.org/10.3847%2F1538-4365%2Fab53d2
Farge, M., Pellegrino, G., & Schneider, K. 2001, Phys Rev Lett, 87 (American Physical Society), 054501
Fox, N. J., Velli, M. C., Bale, S. D., et al. 2016, Space Sci Rev, 204, 7
Greco, A., Chuychai, P., Matthaeus, W. H., Servidio, S., & Dmitruk, P. 2008, Geophys Res Lett, 35, L19111
Greco, A., Matthaeus, W. H., Perri, S., et al. 2017, Space Sci Rev, 214, 1
Hada, T., Koga, D., & Yamamoto, E. 2003, Space Sci Rev, 107, 463
Kasper, J. C., Klein, K. G., Lichko, E., et al. 2021, Phys Rev Lett, 127 (American Physical Society), 255101, https://link.aps.org/doi/10.1103/PhysRevLett.127.255101
Koga, D., & Hada, T. 2003, Space Sci Rev, 107, 495
MacBride, B. T., Forman, M. A., & Smith, C. W. 2005, in Solar Wind 11/SOHO 16, Connecting Sun and Heliosphere, ed. B. Fleck, T. H. Zurbuchen, & H. Lacoste, Vol. 592, 613
Martinović, M. M., Đorđević, A. R., Klein, K. G., et al. 2022, J Geophys Res Space Phys, 127 (John Wiley & Sons, Ltd), e2021JA030182
Matthaeus, W. H., & Velli, M. 2011, Space Sci Rev, 160, 145, https://ui.adsabs.harvard.edu/abs/2011SSRv..160..145M
Matthaeus, W. H., Wan, M., Servidio, S., et al. 2015, Philos Trans R Soc Lond Math Phys Eng Sci, 373 (The Royal Society)
McComas, D. J., Velli, M., Lewis, W. S., et al. 2007, Rev Geophys, 45 (John Wiley & Sons, Ltd), https://doi.org/10.1029/2006RG000195
Moncuquet, M., Meyer-Vernet, N., Issautier, K., et al. 2020, Astrophys J Suppl Ser, 246, 44
Osman, K. T., Matthaeus, W. H., Greco, A., & Servidio, S. 2010a, Astrophys J Lett, 727 (American Astronomical Society), L11, https://doi.org/10.1088/2041-8205/727/1/l11
Osman, K. T., Matthaeus, W. H., Greco, A., & Servidio, S. 2010b, Astrophys J Lett, 727 (American Astronomical Society), L11, https://doi.org/10.1088/2041-8205/727/1/l11
Osman, K. T., Matthaeus, W. H., Hnat, B., & Chapman, S. C. 2012, Phys Rev Lett, 108, 261103, https://link.aps.org/doi/10.1103/PhysRevLett.108.261103
Parashar, T. N., Goldstein, M. L., Maruca, B. A., et al. 2020, Astrophys J Suppl Ser, 246 (American Astronomical Society), 58, https://doi.org/10.3847%2F1538-4365%2Fab64e6

Parashar, T. N., & Matthaeus, W. H. 2016, Astrophys J, 832, 57, http://stacks.iop.org/0004-637X/832/i=1/a=57

Pulupa, M., Bale, S. D., Bonnell, J. W., et al. 2017, J Geophys Res Space Phys, 122 (John Wiley & Sons, Ltd), 2836

Qudsi, R. A., Maruca, B. A., Matthaeus, W. H., et al. 2020, Astrophys J Suppl Ser, 246 (American Astronomical Society), 46, https://doi.org/10.3847%2F1538-4365%2Fab5c19

Richardson, J. D., Paularena, K. I., Lazarus, A. J., & Belcher, J. W. 1995, Geophys R Lett, 22, 325

Sioulas, N., Velli, M., Chhiber, R., et al. 2022, Astrophys J, 927 (American Astronomical Society), 140, https://doi.org/10.3847/1538-4357/ac4fc1

Sorriso-Valvo, L., Marino, R., Carbone, V., et al. 2007, Phys Rev Lett, 99 (American Physical Society), 115001, https://link.aps.org/doi/10.1103/PhysRevLett.99.115001

Taylor, G. I. 1938, Proc R Soc Lond Ser A, 164, 476

Tessein, J. A., Matthaeus, W. H., Wan, M., et al. 2013, Astrophys J, 776 (American Astronomical Society), L8

Tsurutani, B. T., & Smith, E. J. 1979, J Geophys Res Space Phys, 84 (John Wiley & Sons, Ltd), 2773

Veltri, P., & Mangeney, A. 1999, AIP Conf Proc, 471 (American Institute of Physics), 543

Yang, Y., Matthaeus, W. H., Shi, Y., Wan, M., & Chen, S. 2017, Phys Fluids, 29 (American Institute of Physics), 035105

Yordanova, E., Vörös, Z., Sorriso-Valvo, L., Dimmock, A. P., & Kilpua, E. 2021, Astrophys J, 923 (American Astronomical Society), 282, https://doi.org/10.3847/1538-4357/ac4012

Zank, G. P., Zhao, L.-L., Adhikari, L., et al. 2022, Astrophys J Lett, 926 (American Astronomical Society), L16, https://doi.org/10.3847/2041-8213/ac51da